\documentstyle[aps,preprint,epsfig,12pt]{revtex}

\begin{document}


\preprint{TU-630}

\title{
Bottom-Tau Unification in SUSY SU(5) GUT and 
Constraints from $b \rightarrow s \gamma$
and Muon $g-2$
}

\author{Shinji Komine\footnote{e-mail: komine@tuhep.phys.tohoku.ac.jp}
and
Masahiro Yamaguchi\footnote{e-mail: yamaguchi@phys.tohoku.ac.jp}}

\address{Department of Physics, Tohoku University,
Sendai 980-8578, Japan}
\date{\today}
\maketitle

\begin{abstract}
  An analysis is made on bottom-tau Yukawa unification in
  supersymmetric (SUSY) SU(5) grand unified theory (GUT) in the
  framework of minimal supergravity, in
  which the parameter space is restricted by some experimental
  constraints including $\mbox{Br}(b \rightarrow s \gamma)$ and muon
  $g-2$. The bottom-tau unification can be  accommodated to  the measured 
  branching ratio $\mbox{Br}(b \rightarrow s \gamma)$ 
  if superparticle masses are relatively heavy and
  higgsino mass parameter $\mu$ is negative.  On the other hand, if we take the
  latest muon $g-2$ data to require positive SUSY contributions, then
  {\em wrong-sign} threshold corrections at SUSY scale upset the Yukawa 
  unification with more than 20 percent discrepancy. It has to be 
  compensated by superheavy threshold corrections around the GUT scale, which
  constrains models of flavor in SUSY GUT.
  A pattern of the superparticle masses preferred by the three requirements is 
  also commented.
\end{abstract}

\clearpage

\section{Introduction}

Unification of interactions of elementary particles \cite{GerogiGlashow} is a 
very appealing idea which has attracted much attention. 
Supersymmetric (SUSY) SU(5) Grand
Unified Theory (GUT) \cite{SUSY-SU5} is a prototype of such, in which
the standard model gauge group is unified into a SU(5) group, and
quarks and leptons belong to $10$ and $\bar{5}$ representations.

Gauge coupling unification is the most renowned and tested prediction
of the SUSY GUT \cite{CouplingUnification80s}. 
In fact renormalization group analyses 
show that the unification indeed occurs up to
less than a few $\%$ errors \cite{CouplingUnificationLEP}.  
This means that one needs this amount of
corrections coming from physics around the GUT scale and/or above, and
it can easily be accounted for by superheavy thresholds around the GUT
scale, which may include possible SU(5)-violating non-renormalizable
contributions to the gauge coupling constants \cite{HallSarid92}. 
In this sense, the gauge coupling unification is very successful in the 
SUSY GUT scenario.

The grand unification generically has  yet another prediction, {\it i.e.} 
the bottom-tau Yukawa unification\cite{BottomTauNonSUSY,BottomTauSUSY,tanb1or50,Hall-Rattazzi-Sarid,BottomTauAfterHRS}.
There is some
expectation that the bottom quark and the tau lepton which belong to
the same SU(5) multiplet have a unified Yukawa coupling at the GUT
scale.  Apparent difference between the two Yukawa couplings are then
due to renormalization group effects from that scale down to low
energy. In addition, it is known that finite radiative corrections
involving superparticles in the minimal SUSY standard model (MSSM) 
give important contribution to the bottom Yukawa coupling 
\cite{Hall-Rattazzi-Sarid}. The sign and 
magnitude of the threshold corrections at the SUSY scale strongly depend on 
superparticle masses. On the other hand, the superparticle masses now 
suffer from several experimental constraints. They include mass 
bounds on the superparticles and Higgs boson from direct searches, an 
inclusive decay rate of $b \rightarrow s \gamma$,  and
recent measurement of muon $g-2$ or muon anomalous magnetic 
moment $a_{\mu}\equiv (g-2)_{\mu}/2$. 

In this paper, we shall study whether the Yukawa unification occurs
under these experimental constraints.  We perform a numerical analysis
based on minimal supergravity (mSUGRA) and its slight modifications, but most
of our results apply to other cases.  We will clarify how the Yukawa
coupling unification depends on the superparticle mass spectrum and
how it relates to the experimental constraints.  We will show that for
positive $\mu$-parameter where SUSY contribution to muon $g-2$ is
positive, threshold corrections at the SUSY scale make the Yukawa
coupling of bottom quark more than 20 percentage smaller than that of
tau.  On the other hand, for negative $\mu$-parameter, the SUSY
threshold corrections help achieve the bottom-tau unification, while
the $\mbox{Br}(b \rightarrow s \gamma)$ bound is satisfied for
relatively heavy superparticles.  This case may, however, be
confronted with the latest muon $g-2$ data, if its apparent deviation
from the standard model prediction requires positive SUSY
contribution.

Our analysis will provide a useful guide for building models of
flavor.  Deviation from the Yukawa unification, if any, has to be
compensated by corrections at high energy scale around the
GUT scale, if one insists the view of grand unification. The high energy
corrections will be model dependent and thus the issue of the
Yukawa unification will help discriminate models of flavor in SUSY
GUT.

Recently, similar analyses were presented in Refs \cite{SO10Analyses}.
Their concern is the top-bottom-tau Yukawa unification in SO(10) GUT and 
thus they focused on $\tan \beta \gtrsim 50$.
In this parer, we investigate wider regions of the parameter space 
since we are interested in general SUSY GUT models. 

The organization of the paper is the following. In section 2, we will 
briefly review some issues on the bottom-tau unification. In section 3,
we will explain the experimental constraints used in our analysis. Our
numerical analysis is given in section 4. We will summarize our
results in section 5.

\section{bottom-tau Yukawa unification}

In this section, we review some of the important issues on the bottom-tau 
Yukawa unification. In the SU(5) GUT, the SU(2) doublet bottom and the
singlet tau belong to $\Phi(10)$ and the singlet bottom and
the doublet tau (combined with tau neutrino) belong to 
$\Psi(\bar{5})$. If the Yukawa coupling comes from the Higgs 
$\bar{H}(\bar{5})$ as
\begin{equation}
  {\cal L}= y \Phi(10)\Psi(\bar{5})\bar{H}(\bar{5})
\end{equation}
then the bottom and tau have a unified Yukawa coupling $y$. It is
the one given at the GUT scale, and renormalization group (RG) effects
mainly due to QCD make the bottom Yukawa coupling much larger than the
other one. In RG, large Yukawa couplings of top and bottom quarks play an
important role to somewhat lower the bottom Yukawa coupling. It was recognized
some time ago, the QCD RG effect is too strong, predicting a too large
bottom quark mass, unless either of the Yukawa couplings are very
large, {\i.e.} $\tan \beta$ is close to unity, or very large $\gtrsim 50$
\cite{tanb1or50}.  
Here $\tan \beta \equiv \langle H_2 \rangle / \langle H_1
\rangle$ is the ratio of the vacuum expectation values (VEVs) of the two
Higgs bosons $H_1$ and $H_2$. It was also realized that there exist
large threshold corrections at the SUSY scale \cite{Hall-Rattazzi-Sarid}, 
namely finite radiative corrections involving superparticles in the MSSM 
give significant contributions to the bottom mass. 
This is because in the SUSY limit
only one of the Higgses $H_1$ has Yukawa coupling to the bottom quarks
while SUSY breaking allows a new coupling of the other
Higgs $H_2$ to the bottom quarks.  The threshold corrections can be
enhanced especially for large $\tan \beta$. They include
two main contributions: one from a gluino loop and the other from a
chargino loop. They are approximately written as \cite{Hall-Rattazzi-Sarid}
\begin{eqnarray}
   \frac{\delta m_b^{\tilde g}}{m_b} &\approx & 
      \frac{2 \alpha_3}{3 \pi} M_3 \mu 
       I(m_{\tilde b_1}^2,m_{\tilde b_2}^2,M_3^2) \tan \beta,
\\
   \frac{\delta m_b^{\tilde \chi^{\pm}}}{m_b} & \approx &
       - \frac{y_t^2}{16 \pi^2} \mu A_t 
      I(m_{\tilde t_1}^2,m_{\tilde t_2}^2, \mu^2) \tan \beta,
\end{eqnarray}
where $\alpha_3$ is the strong coupling constant, $y_t$ is the top Yukawa
coupling, $M_3$ is the gluino mass, $\mu$ is the supersymmetric Higgsino
mass, $A_t$ is the trilinear SUSY-breaking coupling of the stops to $H_2$,
$m_{\tilde b_{1,2}}$, $m_{\tilde t_{1,2}}$ are sbottom and stop 
mass eigenvalues, respectively. The function $I$ is given
\begin{equation}
      I(x,y,z) =\int^{\infty}_0 \frac{u du}{(u+x)(u+y)(u+z)},
\end{equation}
which is characterized by $I(x,x,0)=1/x$, $I(x,x,x)=1/2x$. 

Of the two threshold corrections the gluino loop  dominates generically
unless the trilinear coupling $A_t$ is very large. Note that minimal
supergravity (mSUGRA) boundary conditions give roughly
\begin{equation}
    A_t \approx 0.2 A_0+2 M_0
\label{eq:At}
\end{equation}
where $A_0$ is the universal trilinear coupling given
at the GUT scale and $M_0$ is the universal gaugino
mass at the GUT scale. The small coefficient in the first term is due
to that the top Yukawa coupling is close to its infra-red fixed point.
Eq.~(\ref{eq:At}) implies that the trilinear stop coupling $A_t$ is mostly
controlled by the universal gaugino mass and thus related to the gluino mass
as far as $A_0$ and $M_0$ are in the same order of magnitude.

In this generic case, the sign of the SUSY threshold corrections 
is solely determined by the sign of $\mu M_3$. Recalling that the
RG effects tend to give too large bottom quark mass, we find the naive 
Yukawa unification at the GUT scale requires this sign to be negative.
This 
argument given here is rather general,
though  we will focus on  the mSUGRA and its variants in our numerical 
analysis. 

Another important point to the SUSY threshold corrections is that they
are non-decoupling effects.  In fact taking all superparticle masses
infinity gives a finite contribution. We will see the importance of
this property later on.

We use the $\overline{MS}$ mass for the bottom quark in the following range
\cite{PDG,LatticeMb} 
\begin{equation}
     m_b^{\overline{MS}}(m_b) = 4.2 \pm 0.2 \quad \mbox{GeV}.
\label{eq:bottom-mass}
\end{equation}
Note that recent analysis based on perturbative 
QCD \cite{Sumino} yields an  estimate of the bottom mass with a much smaller
error.  At this moment, however, other errors coming mainly from the
uncertainty of the top quark mass have comparable effects as
Eq.~(\ref{eq:bottom-mass}), so that reducing the error of the bottom mass
alone does not reduce the errors in the estimate  of the bottom-tau 
 unification. It is 
important to note, however, that the pinpoint evaluation of the bottom
quark mass will become important when future experiments reduce the error
of the top mass.

\section{Experimental Constraints}

Here we summarize experimental constraints used in our analysis.

Null results of superparticle searches and Higgs boson searches give
significant constraints on the allowed parameter space. As for the mass 
bounds of superparticles, we impose $m_{\tilde{\chi}^-_1} > 100 {\rm GeV}$
and $m_{\tilde \tau^-_1} > 80 {\rm GeV}$.
The Higgs mass bound $m_{h} > 113.5$ GeV from the LEP II \cite{LEP2Higgs} 
is also important,
which constrains  $\tan \beta$ and the scale of superparticle
masses.

Another important constraint comes from $b \rightarrow s \gamma$. The
inclusive branching rate Br$(b \rightarrow s \gamma)$ is estimated from
experimental measurements as $(3.21 \pm 0.43 \pm 0.27 ^{+0.18}_{-0.10}) 
\times 10^{-4}$, where the errors are of statistical, systematic and 
theoretical, respectively \cite{CLEO2001}.  
It is known that besides the standard model contribution which is
consistent with the experimental value, SUSY gives two additional
contributions, one from a charged Higgs loop, the other from
superparticle loops. The former always gives an additive contribution to
the standard model. The latter is dominated by a chargino/stop loop, which
can have either sign, depending on the sign of $A_t \mu$. It
turns out that the new contributions tend to cancel for $A_t \mu
\propto M_3 \mu >0$. This case is preferred if we recall that the
standard model contribution already explains the experimental data. On
the other hand, when $M_3 \mu <0$ both the two  contributions have the
same sign as  the standard model contribution. To survive the
experimental constraint, the superparticle mass scale should be large
enough, so that all the additional contributions decouple. Note that this 
decoupling behavior is contrasted to the case of the bottom quark mass in 
which the SUSY contribution remains finite in the decoupling limit.

In the following, we conservatively take the allowed range for Br$(b
\rightarrow s \gamma)$ to be
\begin{equation}
         2 \times 10^{-4} \leq 
         \mbox{Br}(b \rightarrow s \gamma)
        \leq 4.5 \times 10^{-4},
\end{equation}
to constrain the parameter space.\footnote{
We should keep in mind that the $b \rightarrow s \gamma$ process
changes flavor, and thus non-trivial generation mixing in squark
masses coming from unknown flavor physics may change our estimate of
the branching ratio.}  

The other constraint we consider in this analysis is  $a_{\mu}$.
The latest result of the E821 experiment at Brookhaven National Laboratory
reported a possible deviation from the standard model by 2.6-$\sigma$ 
\cite{E821}:
\begin{equation}
   a_{\mu}(E821)-a_{\mu}(\mbox{SM}) =43(16) \times 10^{-10}.
\end{equation}
This may be a signal of new physics beyond the standard model. 
If it is all accounted for by supersymmetry, the SUSY contribution 
$a_{\mu}(\mbox{SUSY})$ should be positive, and of order $10^{-9}$
\cite{g-2afterE821}.
In particular the positiveness requires $M_2 \mu>0$. 
In our analysis, we will include it as an possibly important 
constraint, though  it may be premature to conclude that the latest data 
is a clear evidence of new physics, when one takes into account statistical
significance and uncertainties in the evaluation of the standard model
contributions.  

\section{Analysis}
In this section, we would like to present our numerical results in 
the mSUGRA case.

\subsection{Procedure}
What we will do is to compute the Yukawa couplings of the bottom quark and
the tau lepton at the GUT scale, by using their values at low energy scale to
extrapolate them at the the GUT scale following renormalization group flow.

We take $\alpha_3^{\overline{MS}}(M_Z) = 0.118$, and include SUSY threshold 
effects to obtain $\alpha_3^{\overline{DR}}(M_Z)$ 
\cite{Pierce-Bagger-Mathev-Zhang}.

We fix $\overline{DR}$ Yukawa couplings at $M_Z$ scale as follows.
Having fixed $m_b^{\overline{MS}}(m_b)$, we evolve the running mass from 
$m_b$ to $M_Z$ using the three-loop $\overline{MS}$ RGE's for 
the Standard Model.
Next we convert $m_b$ to the $\overline{DR}$ scheme using a one-loop 
correction factor \cite{MartinVaughn}.
Then we include SUSY threshold effects explained before to extract
the bottom Yukawa coupling in supersymmetric limit. 
As for the tau Yukawa coupling, 
we use $m_{\tau}^{\overline{MS}}(M_Z) = 1.746$ GeV which is obtained 
from the pole mass with two-loop QED corrections \cite{taumass}.
For the top Yukawa, we take the pole mass $m_t^{pole} = 175 {\rm GeV}$. 
We will discuss how change of the top Yukawa affects our results later on.

To obtain the gauge and Yukawa couplings at the unification scale, 
we use two-loop level RGE's for the minimal supersymmetric standard model 
(MSSM) \cite{2loopRGEs}.
We define $M_{\rm GUT}$ to be a point where $\alpha_1$ and $\alpha_2$ meet.
\footnote{At the $M_{\rm GUT}$, $\alpha_3$ is smaller than $\alpha_1$ and 
$\alpha_2$ by up to 4 \%. We do not care this small discrepancy here.}
Here it is convenient to define $\Delta_{b-\tau} \equiv 
(Y_b(M_{GUT}) -Y_{\tau}(M_{GUT}))/Y_b(M_{GUT})$ to characterize how well
the Yukawa unification achieves.

To evaluate the SUSY threshold corrections, the Higgs mass, 
Br$(b \to s \gamma)$ and  SUSY contribution to $a_{\mu}$, 
we need to know the SUSY breaking masses at the weak scale. 
In our analysis, these are computed by using one-loop level RGE's for the MSSM.
(For the gauge and Yukawa couplings, we use two-loop level RGE's.)
We evaluate the parameter $\mu$, the masses of the charged Higgs 
and the pseudoscalar Higgs at the energy scale of a geometrical mean of stop 
masses $\sqrt{ m_{\tilde{t_1}} m_{\tilde{t_2}} }$ 
with one-loop effective potential.
We calculate the lightest Higgs boson mass using an approximate formula 
including two-loop corrections,\footnote{
This approximate result can be different from the exact one 
by up to 2 GeV \cite{Heinemeyer:2000nz}. 
The lower limit on the SUSY scale from the Higgs boson mass 
constraint changes by a few tens GeV because of this uncertainty. }
{\it FeynHiggsFast} \cite{Heinemeyer:2000nz}.
Since the bottom Yukawa coupling receive the large SUSY threshold, 
we have to take this effect into account to calculate 
Br$(b \rightarrow s \gamma)$ in the large $\tan \beta$ case.
We follow the NLO formalism
that can be applicable to large $\tan \beta$ case \cite{bsgamma}.
And for the SUSY contribution $\Delta a_{\mu}$ to the muon anomalous moment,
we use the formula in the one loop approximation\cite{gminus2}.
  
\subsection{Supergravity Boundary Conditions}
In our analysis, we give the boundary conditions of the soft SUSY breaking
masses at the GUT scale, and discard possibly large contributions above it
\cite{KawamuraMurayamaYamaguchi}. 
At the GUT scale, the scalar masses of 10, $\bar 5$ representations of 
SU(5) $m^2_{\phi}$, $m^2_{\psi}$, respectively, and the Higgs boson masses are
assumed to be in the form
\begin{eqnarray}
  m^2_{\psi}(M_{GUT}) = m^2_{\psi 0} {\bf 1} \quad,
  m^2_{\phi}(M_{GUT}) = m^2_{\phi 0} {\bf 1} \quad,
  m^2_{H_1}(M_{GUT}) = m^2_{H_1 0} \quad,
  m^2_{H_2}(M_{GUT}) = m^2_{H_2 0}.
\end{eqnarray}
As for trilinear couplings, we take the following form,
\begin{eqnarray}
  A_u(M_{GUT}) = A_d(M_{GUT}) = A_e(M_{GUT}) = A_0 {\bf 1}.
\end{eqnarray}
Three gaugino masses in the MSSM are unified to $M_0$ at the GUT scale.
We take a convention that $M_0 > 0$.
By solving the RGE's and imposing the radiative electroweak symmetry breaking, 
one obtains all SUSY parameters at the weak scale. 

For later convenience, we give approximate formulas of the Higgsino mass $\mu$,
scalar masses in the third generation and the charged Higgs mass since 
SUSY contribution to the muon $g-2$ $\Delta a_{\mu}$, 
$b \rightarrow s \gamma$ and the lightest Higgs mass depend on these masses.
For $\tan \beta \lesssim 10$ where effects of the bottom and tau Yukawa 
couplings to RG running of scalar masses are neglected, one finds, 
\begin{eqnarray}
  \mu^2 &\simeq& -0.64 m^2_{H_2 0}  +0.72 m^2_{\phi 0}  +2.7 M_0^2  
  +0.42 M_0 A_0,
  \\
  m^2_{H^-} &\simeq& m^2_{H_1 0} -0.64 m^2_{H_2 0} +0.72 m^2_{\phi 0} 
  +3.2 M_0^2 
  +0.42 M_0 A_0, \\
  m^2_{\tilde{t}_L} &\simeq& 0.76 m^2_{\phi 0} -0.12 m^2_{H_2} +5.5 M_0^2
  -0.14 M_0 A_0, \\
  m^2_{\tilde{t}_R} &\simeq& 0.52 m^2_{\phi 0} -0.24 m^2_{H_2} +3.9 M_0^2
  -0.28 M_0 A_0 .
\end{eqnarray}
Sbottom and stau masses are the same as the first two generations.
In the mSUGRA case where all scalar masses are taken to be  
universal,
$m^2_{\psi 0} = m^2_{\phi 0} = m^2_{H_1 0} = m^2_{H_2 0} = m^2_0$,
we find 
\begin{eqnarray}
  \mu^2 &\simeq& 0.08 m^2_0 +2.7 M_0^2 +0.42 M_0 A_0, \\
  m^2_{H^-} &\simeq& 0.36 m^2_0 +3.2 M_0^2 +0.42 M_0 A_0, \\
  m^2_{\tilde{t}_L} &\simeq& 0.64 m^2_0 +5.5 M_0^2 -0.14 M_0 A_0, \\
  m^2_{\tilde{t}_R} &\simeq& 0.28 m^2_0 +3.9 M_0^2 -0.28 M_0 A_0.
\end{eqnarray}
These equations show that coefficients in front of $m_0^2$ get small,
especially for $\mu^2$, because of top Yukawa effects.
Hence in the mSUGRA case these values are mostly determined by the 
universal gaugino mass, and larger than the Bino mass $M_1 \simeq 0.4 M_0$ 
or Wino mass $M_2 \simeq 0.8 M_0$. 

\subsection{Results}
Here we present our numerical results of the bottom-tau unification,
$\Delta_{b-\tau}=(Y_b(M_{GUT}) -Y_{\tau}(M_{GUT}))/Y_b(M_{GUT})$ under
the constraints from the Higgs mass, $b \rightarrow s \gamma$, and
$a_{\mu}$ in the mSUGRA case.

Firstly we would like to discuss how the bottom-tau unification as
well as the experimental constraints are sensitive to the sign of the
$\mu$ parameter. For illustration, we take $\tan \beta=30$, $A_0=0$.

In Fig. 1, we draw contours on the bottom-tau unification for negative $\mu$.
We find that the unification achieves within 5 percent in the large region
of the parameter space.  Notice that the present uncertainties of the bottom 
mass and the top quark mass will generate the error of about 5 percent,
and also we expect that a few percentage superheavy thresholds are easily
obtained. Thus in this case, we conclude that the bottom-tau unification
is successful.

A large part of the parameter space is, however, eliminated by
the experimental constraints. The Higgs mass bound is satisfied when
the gaugino mass $M_0$ is larger than about 300 GeV, while the 
requirement from Br$(b \rightarrow s \gamma)$ is a bit severer. In fact
it requires relatively heavy superparticles, for example, $M_0 \gtrsim 700$
GeV for small scalar mass, or $m_0 \gtrsim 1,200$ GeV for small gaugino 
mass.  This is because, in the negative $\mu$ case, the SUSY contribution
as well as the charged Higgs contribution to the $b \rightarrow s \gamma$
process are additive to the standard model, and thus these contributions
have to decouple in order to be consistent with the experimental data.
Note that even with such heavy superparticle masses the bottom-tau
unification can be explained as one sees in Fig. 1. 

On the other hand, the SUSY contribution to the muon anomalous
magnetic moment is always negative in this case, which is disfavored
by the data from the E821 experiment. Thus if $a_{\mu}$ requires a
positive SUSY contribution, then the $\mu <0$ case is ruled out.  More
conservatively if one allows, for instance, 3-$\sigma$ deviation from
the central value of their result, the $\mu <0$ case is allowed. In
Fig. 1, we plot a contour of the SUSY contribution $\Delta a_{\mu}=-5
\times 10^{-10}$, corresponding to the 3-$\sigma$ deviation. This gives
a much severer constraint than $b \rightarrow s \gamma$, and the
bottom-tau unification (within 5 \%) is achieved only for very heavy
superparticle mass spectrum in which sleptons , for instance, weigh
much more than 1 TeV.

Next we will consider the positive $\mu$ case. Compared to the negative
$\mu$ case, the Higgs mass bound gives more or less a similar constraint,
while the constraint from $b \rightarrow s \gamma$ becomes much weaker. 
This is because the new contributions tend to cancel each other, and thus
a wider region gets allowed than the previous case. The SUSY contribution
to $a_{\mu}$ is positive, which can easily be accommodated to the
E821 data. Now the bottom-tau unification becomes in bad shape, namely
the exploration of the Yukawa couplings from the low energy data to
the GUT scale using the RGE's with the SUSY threshold included yields
discrepancy of typically more than 20 percent. In fact for $\mu>0$
the SUSY threshold to the bottom quark mass is always positive, which
makes the bottom Yukawa in the SUSY limit small. This in turn gives a too 
small bottom Yukawa at the GUT scale after the renormalization group flow.

Let us next discuss how the bottom-tau unification depends on  $\tan \beta$. 
To see this, we randomly generated parameters in the range 
$100 {\rm GeV} \leq m_0 \leq 1500 {\rm GeV}$, 
$100 {\rm GeV} \leq M_0 \leq 1500 {\rm GeV}$, and required that
$m_h \geq 113.5 {\rm GeV}$, $m_{\tilde{\chi}_1} \geq 100 {\rm GeV}$, 
$m_{\tilde{\tau}_1} \geq 80 {\rm GeV}$, 
$2 \times 10^{-4} \leq {\rm Br}(b \rightarrow s \gamma) \leq 
4.5 \times 10^{-4}$.
In Fig. 3, we plot parameter points which survive the constraints given above
for the positive $\mu$ case. Points marked with square ($\Box$) 
are those in which  $\Delta a_{\mu}$ is consistent with 
the result of the E821 experiment at the $2 \sigma$ level. 
The upper end of the band corresponds to heavier SUSY scale.
We find the large discrepancy from the naive expectation of the bottom-tau
unification ($\Delta_{b-\tau}=0$),  especially for large $\tan \beta$.
Notice that the points which satisfy the 2-$\sigma$ constraint on $a_{\mu}$
tend to give larger deviation in the bottom-tau unification. This can be
understood in the following way.  Recall that the dominant SUSY contribution
to $a_{\mu}$ comes from sneutrino-chargino loop.
To make the SUSY contribution large, we therefore require that the 
sneutrino is relatively light, which means that in the context of the mSUGRA
the sbottoms and the gluino cannot be very heavy. This results in a 
large SUSY threshold correction, which makes the bottom-tau unification worse.

 The discrepancy should be explained by high energy thresholds, such as
threshold corrections by superheavy GUT-particles and/or effects from
non-renormalizable operators. It is highly model dependent whether we 
can get more than 20 percentage corrections and thus this will provide a
critical test of models of flavor in the SUSY GUT.
   
A similar plot is given for $\mu < 0$ in Fig. 4. 
For $\tan \beta \lesssim 35$, the bottom-tau unification works well 
within 10 \%. 
In this case as we anticipate,
there is no point which give the 2-$\sigma$ $\Delta a_{\mu}$. 
There are, however, points where $\Delta a_{\mu}$ is consistent with 
the E821 result within 3-$\sigma$ level, which are marked with circle 
($\circ$) in Fig. 4.

So far we have discussed the mSUGRA case.
One characteristic feature of the mSUGRA model is that $\mu$ is almost 
determined by $M_0$ and is insensitive to $m_0$, 
as far as $m_0$ and $M_0$ is in the same order of magnitude.
But once we relax the mSUGRA boundary conditions and suppose, for instance,
that 
the Higgs boson masses and the sfermion masses are different at the GUT scale,
then the situation can change.
For example, we consider the case with the following boundary conditions 
\begin{eqnarray}
  m^2_{\psi}(M_G) = m^2_{\phi}(M_G) = m^2_0 {\bf 1}, \quad
  m^2_{H_u}(M_G) = m^2_{H_d}(M_G) = (1.5 m_0)^2.
\end{eqnarray}
Then $\mu$ parameter, the charged Higgs mass and stop masses are 
approximately estimated as 
\begin{eqnarray}
  \mu^2 &\simeq& -0.72 m^2_0 +2.7 M_0^2 +0.42 M_0 A_0, \\
  m_{H^-}^2 &\simeq& 1.5 m^2_0 +3.2 M_0^2 +0.42 M_0 A_0, \\
  m_{\tilde{t}_L}^2 &\simeq& 0.49 m^2_0 +5.5 M_0^2 -0.14 M_0 A_0, \\
  m_{\tilde{t}_R}^2 &\simeq& -0.02 m^2_0 +3.9 M_0^2 -0.28 M_0 A_0
\end{eqnarray}
Comparing to the mSUGRA case, we find that $\mu^2$ can be smaller, thanks
to the negative coefficient in front of $m_0^2$. 
Since the SUSY threshold correction to the bottom mass is proportional 
to $\mu$, it will be suppressed.
The light Higgsino mass, on the other hand, does not suppress $\Delta a_{\mu}$
since a dominant SUSY contribution to it comes from a chargino/sneutrino loop
in which both the Wino and Higgsino propagate.
In addition, the Higgs mass and $b \rightarrow s \gamma$ constraints do 
not change so much since the charged Higgs boson mass and the stop mass mostly
depend on the universal gaugino mass as the mSUGRA case.  
A numerical result for $\mu>0$ is shown in Fig. 5.
We find that the difficulty of the bottom-tau unification for the positive
$\mu$ case is somewhat ameliorated, {\it i.e.} the discrepancy can be
as small as 15 \%.

Finally we wish to comment on uncertainties coming from the bottom mass
and the top mass.  As was discussed in a previous section, 
we conservatively took the error of the bottom mass estimate to be 5\%.
This error almost linearly reflects the uncertainty in the bottom-tau 
unification.
On the other hand, the uncertainty from the top mass is more subtle. We
shifted the pole mass of the top quark of 175 GeV by $\pm 5$ GeV, and found
that the resulting change in the bottom-tau unification is up to 5 \%.
These uncertainties should be kept in mind when discussing this issue. 
\footnote{
Notice that the shift of the top quark mass also affects  the lightest 
Higgs boson mass by about $10 {\rm GeV}$.
Hence for $m_t^{pole} = 180 {\rm GeV}$, 
a lower  $\tan \beta$ will be allowed.}

\section{conclusions and discussion}
In this paper, we have analyzed the bottom-tau unification in the SUSY
SU(5) scenario. The sign of the SUSY threshold corrections plays an
important role. 
It is determined by the sign of the $\mu$ parameter,
in the mSUGRA scenario where the gaugino masses have
the same sign and the SUSY-breaking trilinear coupling of
stop-stop-Higgs $A_t$ is strongly correlated with the sign of the gluino. 
In fact the bottom-tau unification works well for $\mu <0$. On the other hand,
the consistency of the measured branching ratio of $b \rightarrow 
s \gamma$ with the standard model prediction implies that the additional
contributions in the MSSM should be small. This favors the case where
the charged Higgs contribution is partially cancelled with the stop/chargino
loop contribution, which occurs in the positive $\mu$ case. We showed
that the bottom-tau unification is in accord with the
$b \rightarrow s \gamma$ constraint for $\mu<0$ if the superparticle masses
are relatively heavy.  In this case the SUSY contributions to the $b
\rightarrow s \gamma$ process decouples.  Since the SUSY threshold to
the bottom mass does not have this decoupling property, the bottom-tau
unification does not occur in $\mu>0$. 

Inclusion of the measurement of $a_{\mu}$ at the E821 experiment makes
the bottom-tau unification difficult.  The positive SUSY contribution
to $a_{\mu}$ suggested by the experiment favors $\mu>0$, opposite
to the sign preferred by the Yukawa unification.  In fact, we showed that
the discrepancy of the unification is larger than 20 \% for $\mu>0$ in the
mSUGRA. The shift of the soft mass for Higgses does not significantly
improve the situation. Thus we conclude that the bottom-tau unification
in the mSUGRA-type SUSY breaking will conflict with the $a_{\mu}$ measurement,
if the deviation is confirmed in future.

Before closing, we would like to discuss what kind of SUSY breaking pattern
is needed to explain all three, namely 
the bottom-tau unification, $b \rightarrow 
s \gamma$, and the discrepancy of $a_{\mu}$. We infer that the following
superparticle mass spectrum will be satisfactory:
\begin{itemize}
\item The hypothesis of the universal gaugino mass is abandoned, and the 
gluino mass and the Wino mass should have an opposite sign.
\item To obtain a sizable effect to $a_{\mu}$, the sneutrino and
the charginos are relatively light.
\item To suppress the  additional contributions to $b \rightarrow s \gamma$,
we need the heavy charged Higgs and the heavy stop. This may be
realized if the soft mass to the Higgs is large and different from the other
scalar masses, and the gluino is heavy so that the stop acquires mass
via the renormalization group effect.
\end{itemize}

To illustrate the first point, a plot is given in Fig. 6 for the case 
that the sign of the gluino mass is opposite to the Bino and the Wino mass, 
$M_1(M_{GUT}) = M_2(M_{GUT}) = -M_3(M_{GUT})(>0)$,
while all scalar masses are universal at the GUT scale.
In this case the bottom-tau unification can be achieved for $M_3 \mu <0$
while the E821 result can be explained by SUSY effects since $M_2 \mu >0$.
The negative sign of the gluino mass results in negative value of $A_t$,
therefore the $b \rightarrow s \gamma$ constraint becomes severer.
However, as shown in Fig. 6,  we find a region where all constraints 
are satisfied, in which the muon $g-2$ is consistent with the E821 result 
at 2-$\sigma$ level, and the bottom-tau unification is achieved within 10 \%. 
If we impose 1-$\sigma$ constraint on $a_{\mu}$, however, 
the allowed region will disappear.
It is interesting to note that future improvement of the $a_{\mu}$ measurement
may be able to eliminate the allowed region \cite{KomineMoroiYamaguchi}.

We expect that this type of the soft masses may be provided in some
classes of mediation mechanisms of SUSY breaking, including the case
where SUSY is broken in an SU(5) violating 3-brane and thus non-universal 
gaugino masses may arise \cite{Hall-Nomura}.  
Further discussion along this line should be encouraged in particular when
the deviation in $a_{\mu}$ is confirmed in future.

\section*{Acknowledgment}               
We would like to thank M. Bando, N. Maekawa, T. Moroi, Y. Sumino 
Y. Yamada and K. Yoshioka for valuable discussions. One of us (MY)
thanks the Summer Institute 2001 held in Yamanashi, Japan. 
This work was supported in part by the
Grant-in-aid from the Ministry of Education, Culture, Sports, Science
and Technology, Japan, priority area (\#707) ``Supersymmetry and
unified theory of elementary particles,'' and in part by the
Grants-in-aid No.11640246 and No.12047201.

\begin{figure}
\begin{center}
\leavevmode 
\psfig{file=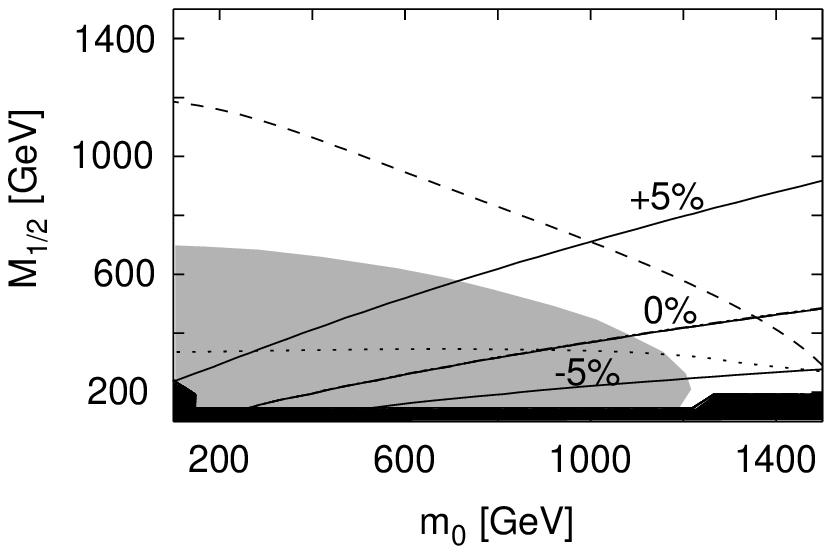,width=10cm}
\end{center}
\caption{The solid lines are coutours of $\Delta_{b-\tau}$ on $m_0$ v.s 
$M_{1/2}$ plane for $\tan \beta = 30$ and $\mu <0$ 
with mSUGRA boundary condition.
The lightly shaded region is the parameter space where 
Br$(b \rightarrow s \gamma) > 4.5 \times 10^{-4}$ and thus excluded by the 
experimental results. $m_h = 113.5$ GeV line is shown as the dotted line.
The region above the dashed line is consistent with the E821 result
at 3-$\sigma$ level, i.e.,
$\Delta a_{\mu} > - 5 \times 10^{-10}$.
Electroweak symmetry breaking does not occur correctly in the darkly-shaded 
region. }
\label{fig1}
\end{figure}

\begin{figure}
\begin{center}
\leavevmode 
\psfig{file=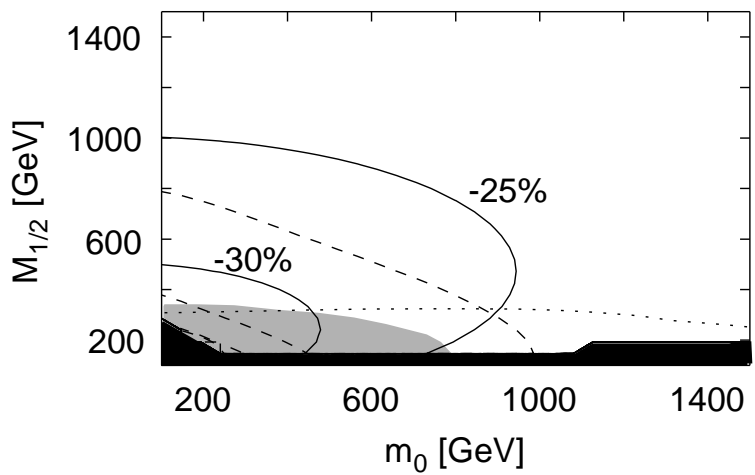,width=10cm}
\end{center}
\caption{Contours of $\Delta_{b-\tau}$ for $\tan \beta = 30$ and $\mu >0$ .
In the lightly-shaded region Br$(b \to s \gamma) < 2 \times 10^{-4}$, thus 
this region is excluded.
The dashed lines are for $\Delta a_{\mu} = 11 \times 10^{-10}$, 
$\Delta a_{\mu} = 43 \times 10^{-10}$ and $\Delta a_{\mu}= 75 \times 10^{-10}$
from the bottom-left.}
\end{figure}

\begin{figure}
\begin{center}
\leavevmode 
\psfig{file=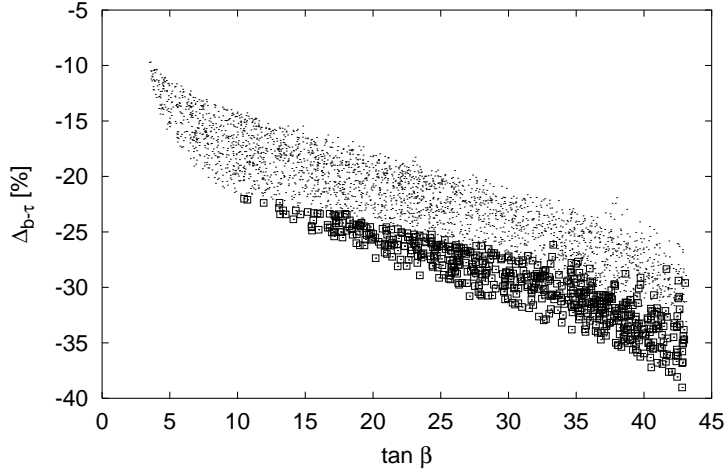,width=10cm}
\end{center}
\caption{Plots of $\Delta_{b-\tau}$ versus $\tan \beta$ for the positive 
$\mu$ case with mSUGRA boundary condition. 
All points are consistent with the lower limit of the lightest 
Higgs mass, chargino mass, stau mass, and the $b \rightarrow s \gamma$ result.
Points marked with square are consistent with the E821 result at 2-$\sigma$ 
level. }
\end{figure}

\begin{figure}
\begin{center}
\leavevmode 
\psfig{file=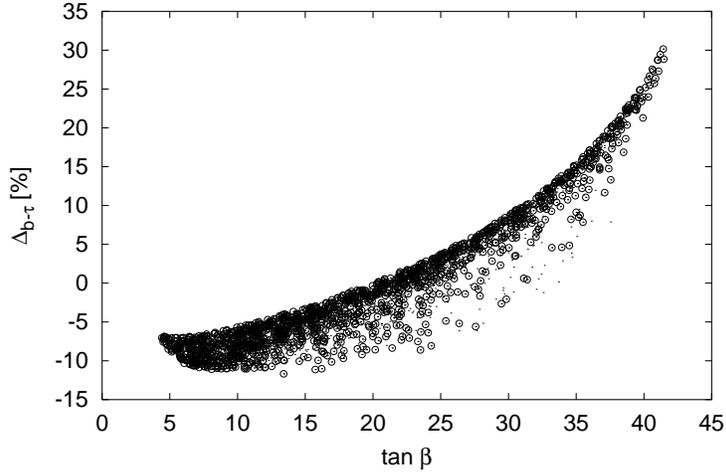,width=10cm}
\end{center}
\caption{Same plot as the Fig. 3, for the negative $\mu$ case. In this figure,
points marked with circle are consistent with the E821 result at 3-$\sigma$ 
level.}
\end{figure}

\begin{figure}
\begin{center}
\leavevmode 
\psfig{file=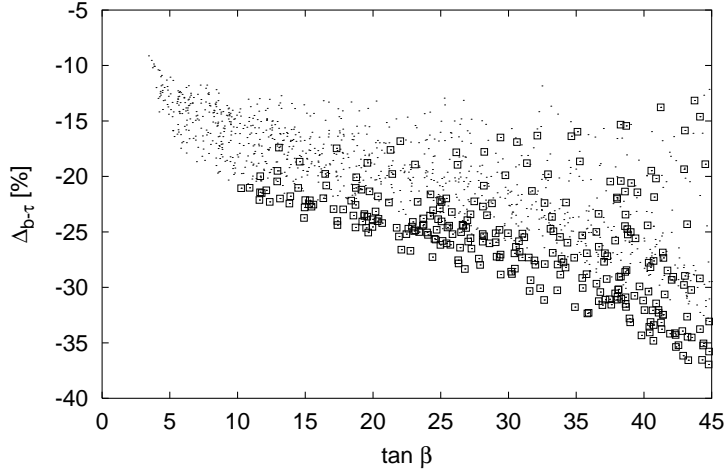,width=10cm}
\end{center}
\caption{Same as Fig. 3 except that $m^2_{H_1} = m^2_{H_2} = (1.5 m_0)^2$, 
$m^2_{\phi} = m^2_{\psi} = m^2_0$ at the GUT scale. The plot is for $\mu > 0$.}
\end{figure}

\begin{figure}
\begin{center}
\leavevmode 
\psfig{file=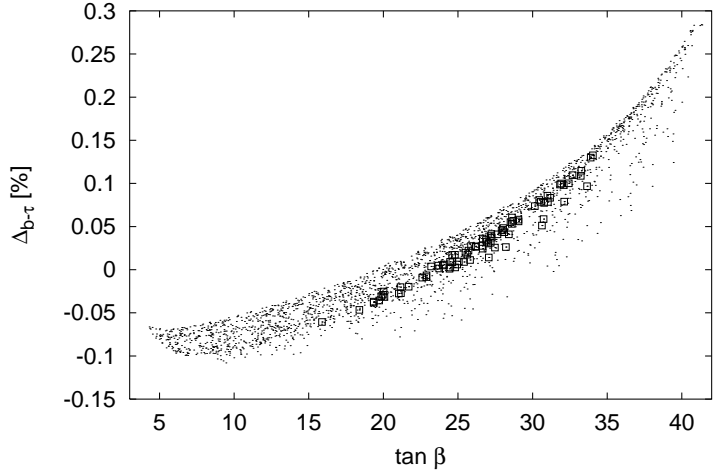,width=10cm}
\end{center}
\caption{Same as Fig. 3 except that 
$M_1(M_{GUT}) = M_2(M_{GUT}) = - M_3(M_{GUT})$
at the GUT scale. The soft scalar masses are assumed to be
universal at the GUT scale  and  $\mu>0$.  }
\end{figure}

\end{document}